\documentclass[twocolumn,aps,floats,floatfix,nofootinbib,prd,superscriptaddress,tightenlines]{revtex4-1}

\usepackage{amsmath,amsfonts,amssymb}

\newcommand{\Op}[2]{\mathcal{O}_{#1}(\eta_{#2})}

\newcommand{\ee}[3]{(\eta_{#1}\cdot\eta_{#2})^{#3}}
\newcommand{\e}[3]{\eta_{#1}^{#2_{#3}}}
\newcommand{\D}{\mathcal{D}}
\newcommand{\A}{\mathcal{A}}
\newcommand{\cOPE}[4]{{}_{#1}c_{#2#3}^{\phantom{#2#3}#4}}
\newcommand{\DOPE}[4]{{}_{#1}\D_{#2#3}^{\phantom{#2#3}#4}}
\newcommand{\tOPE}[6]{{}_{#1}t_{#2#3}^{#5#6#4}}

\begin{document}

\title{A recipe for conformal blocks}

\author{Jean-Fran\c{c}ois Fortin}
\affiliation{D\'epartement de Physique, de G\'enie Physique et d'Optique\\Universit\'e Laval, Qu\'ebec, QC G1V 0A6, Canada}

\author{Witold Skiba}
\affiliation{Department of Physics, Yale University, New Haven, CT 06520, USA}

\begin{abstract}

We describe a prescription for constructing conformal blocks in conformal field theories in any space-time dimension with arbitrary quantum numbers. Our procedure reduces the calculation of conformal blocks to constructing certain group theoretic structures that depend on the quantum numbers of primary operators. These structures project into irreducible Lorentz representations. Once the Lorentz quantum numbers are accounted for there are no further calculations left to do. We compute a multivariable generalization of the Exton function. This generalized Exton function, together with the group theoretic structures, can be used to construct conformal blocks for four-point as well as higher-point correlation functions.

\end{abstract}

\maketitle

Exact results in strongly-coupled relativistic field theories are scarce. Analytic treatment is sometimes possible due to additional symmetries extending Poincar\'e invariance. There are well-known solutions exploiting either conformal symmetry or supersymmetry. Examples are conformal theories in two dimensions and the minimal models in which the scaling dimensions are calculable~\cite{minimal}. Other examples are $\mathcal{N}=2$ supersymmetric theories in four dimensions in which the low-energy gauge coupling is calculable~\cite{SeibergWitten}. In both of these cases, additional symmetries constrain the possible forms of interactions and given such constraints various consistency conditions are sufficient to obtain analytic solutions.

Conformal symmetry in any number of dimensions restricts the form of two- and three-point correlation functions of primary operators leaving only a finite number of unspecified numerical constants. Four-point and higher-point functions depend on conformally invariant combinations of variables, the conformal cross-ratios. A surprising aspect of conformal field theories (CFTs) is that symmetries also constrain correlation functions beyond three points despite the existence of the invariant cross-ratios for four, or more, coordinates. What differentiates conformal symmetry restrictions on three-point correlation functions from the higher ones is that the quantum numbers of the operators are sufficient to determine the form of the correlator. With four, or more, points one needs to specify not only the quantum numbers of the operators at each point, the ``external" operators, but additional quantum numbers of ``exchange" operators that do not appear explicitly in the correlator. The functional form of four-point, or higher-point, correlators with given external and exchange quantum numbers are known as the conformal blocks.

Conformal blocks are the main inputs for non-perturbative studies of conformal field theories. The bootstrap approach to conformal field theories invented in the seventies~\cite{bootstrap} relies on very minimal assumptions such as crossing symmetry and unitarity. A correlation function can be expressed in several ways in terms of the conformal blocks. Since different ways of calculating the same quantity must be equivalent one obtains constraints on the scaling dimensions of the operators and the three-point function coefficients. In some cases, like the minimal models in two dimensions~\cite{minimal}, the bootstrap turns out to be powerful enough to completely determine all parameters. For theories in more two dimensions, there has been a lot of progress in the last decade mostly in numerical bootstrap leading to many interesting results, some example are in Ref.~\cite{bootstrap4d}.

Beyond two dimensions, the conformal blocks are not known in general. There are several approaches to computing the blocks, but the applicability is often restricted to a particular dimension of space-time or to a particular set of quantum numbers of the operators~\cite{blocks-examples1,blocks-examples2,blocks-examples3}. Here, we present a unified treatment that yields any conformal block completing the approach we outlined in~\cite{outline}.

We utilize two standard techniques. First, the embedding space in which the $d$-dimensional space with coordinates $x^\mu$ is embedded on the light cone of a $(d+2)$-dimensional projective space~\cite{embedding}. We refer to the embedding space coordinates as $\eta^A$ with $\eta_A\eta^A=0$ and the identification $\eta^A\sim\lambda\eta^A$ for $\lambda>0$. The position space coordinate is $x^\mu=\eta^\mu/(-\eta^{d+1}+\eta^{d+2})$. Conformal symmetry acts linearly on the $\eta^A$ coordinates. Second, we successively use the operator product expansion (OPE) inside correlation functions to reduce higher-point functions to lower-point ones. This iterative process is possible because the OPE is convergent inside correlators of conformal theories~\cite{OPE-convergence}.

These two tricks are old and well known. What allowed us to derive completely general results were two further observations. A careful choice of a differential operator on the embedding space made the necessary calculations manageable. All operators are uplifted to the embedding space in the same way, independently of their quantum numbers, using spinor indices only.

In Sec.~\ref{sec:operators}, we describe how operators are uplifted to the embedding space and we outline the ingredients of the OPE. The main algebraic results are presented in Sec.~\ref{sec:Exton} leading to a generalization of the Exton $G$-function~\cite{Exton}. We summarize our approch in Sec.~\ref{sec:discussion}.


\section{Operators, derivatives and the OPE}\label{sec:operators}

Primary operators in CFTs are characterized by their scaling dimensions and Lorentz quantum numbers. We are going to work exclusively in the embedding space, in which the scaling dimension is simply related to the homogeneity degree of an operator
\begin{equation}
\eta^A\frac{\partial}{\partial\eta^A}\mathcal{O}(\eta)=-\tau_\mathcal{O}\mathcal{O}(\eta),
\end{equation}
where $\tau_\mathcal{O}$ is the twist of the operator, which is given in terms of its dimension and spin $\tau_\mathcal{O}=\Delta_\mathcal{O}-S_\mathcal{O}$. For now, we suppressed any information about Lorentz representation of $\mathcal{O}$.

Since the embedding space is larger than the original space, operators with Lorentz indices have additional components that need to be removed. We found it most convenient to represent Lorentz quantum numbers as tensor products of spinors. Any representation of $SO(d+2)$ is contained in a tensor product of the spinor representations. We will not specify the signature of space as it will not play a role here. For simplicity of the presentation, let us assume that $d$ is odd so we do not need to distinguish different spinor representations. This too, is a technical detail that is not relevant. With spinor indices explicit, we will write operators as $\mathcal{O}_{a_1\ldots a_n}(\eta)$.

The following transversality condition
\begin{equation}\label{eq:transverse}
\eta_A(\Gamma^A)_a^{\phantom{a}a_i}\, \mathcal{O}_{a_1\ldots a_i \ldots a_n}(\eta)=0,
\end{equation}
imposed on every index $i=1,\ldots,n$ removes the unwanted components of $\mathcal{O}_{a_1\ldots a_n}(\eta)$. In the equation above, $\Gamma^A$ are the usual Dirac matrices in $d+2$ dimensions. We want the operators to transform in irreducible representations, so we assume that $(\mathcal{P}_{\boldsymbol{N}})_{a_1\ldots a_n}^{\phantom{a_1\cdots a_n} b_1 \ldots b_n} \mathcal{O}_{b_1\ldots b_n} = \mathcal{O}_{a_1\ldots  a_n}$. $\mathcal{P}_{\boldsymbol{N}}$ is a projection operator from the tensor product of $n$ spinors into any irreducible representation, denoted $\boldsymbol{N}$, in that product. Labelling representations by their Dynkin indices, the position space operator with $\boldsymbol{N}_p=(n_1,\ldots,n_r)$ is related to the embedding space representation $\boldsymbol{N}=(0,n_1,\ldots,n_r)$. This way of embedding operators is convenient because the number of spinor indices in the position and embedding spaces are exactly the same. The transversality condition in Eq.~(\ref{eq:transverse}) changes the representation from $\boldsymbol{N}$ to $\boldsymbol{N}_p$.

Schematically, the OPE of two operators can be written as
\begin{equation}\label{eq:OPE-schematic}
\Op{i}{1}\Op{j}{2}=\sum_k\sum_{a=1}^{N_{ijk}}\cOPE{a}{i}{j}{k}\DOPE{a}{i}{j}{k}(\eta_1,\eta_2)\Op{k}{2},
\end{equation}
where the operators on both sides can be in arbitrary Lorentz representations. The sum over $a$ runs over different possibilities for contracting Lorentz indices of a given set of operators. The number of terms in this sum is the number of independent coefficients in the three-point function $\langle\Op{i}{1}\Op{j}{2}\Op{k}{3}\rangle$. The operator on the right-hand side of the OPE is assumed to be at $\eta_2$ although one could make an equivalent choice of $\eta_1$ instead. This operator has to be on the null light cone, so a symmetric choice that treats both coordinates on the same footing is not possible.

The derivative operator $\DOPE{a}{i}{j}{k}$ in Eq.~(\ref{eq:OPE-schematic}) serves two goals. It soaks up some number of Lorentz vector indices to ensure that the OPE is Lorentz covariant. It is also needed to ensure that both sides of the OPE have the same degree of homogeneity with respect to coordinates $\eta_1$ and $\eta_2$.

The derivative operator is not unique as the OPE dictates only the number of vector indices and scaling with respect to the coordinates. One constraint is that the derivatives cannot  take fields defined on the light cone outside such light cone. Our choice is driven by computational convenience. The basic building blocks for the derivatives are, see \cite{derivatives} for details,
\begin{eqnarray}
\D_{12}^A&=&\ee{1}{2}{\frac{1}{2}}\A_{12}^{AB}\partial_{2B},\ \ \D_{12}^2=\D_{12}^A\D_{12A},\\ 
\A_{12}^{AB}&=&\frac{1}{\ee{1}{2}{}}[\ee{1}{2}{}g^{AB}-\e{1}{A}{}\e{2}{B}{}-\e{1}{B}{}\e{2}{A}{}].
\end{eqnarray} 
The transverse metric $\A_{12}^{AB}$ appears in many places in our construction because $\eta_{1A}\A_{12}^{AB}=\eta_{2A}\A_{12}^{AB}=0$. The operator that has really convenient properties is
\begin{equation}
\D_{12|h}^A=\frac{\e{2}{A}{}\D_{12}^2}{\ee{1}{2}{\frac{1}{2}}}+2h\D_{12}^A-h(d+2h-2)\frac{\e{1}{A}{}}{\ee{1}{2}{\frac{1}{2}}},
\end{equation}
satisfying for example $\D_{12|h+1}^A\D_{12|h}^B=\D_{12|h+1}^B\D_{12|h}^A$ and $\D_{12}^{2h}\D_{12|h'}^A=\D_{12|h+h'}^A\D_{12}^{2h}$. For now, the parameter $h$ is arbitrary, but when $\D_{12|h}^A$ appears in the OPE $h$ will be uniquely determined.

One of the most useful identities for the derivatives is
\begin{equation}\label{eq:derivative-coordinates}
\D_{12|h+n}^{A_n}\ldots\D_{12|h+1}^{A_1}\D_{12}^{2h}=\frac{\D_{12}^{2(h+n)}\e{2}{A}{1}\ldots\e{2}{A}{n}}{\ee{1}{2}{\frac{n}{2}}},
\end{equation}
because it allows us to trade the scalar derivative acting on coordinates for derivatives with Lorentz indices. This combination is so useful that we define
\begin{equation}
\D_{12}^{(d,h,n)A_1\ldots A_n}=\D_{12|h+n}^{A_n}\ldots\D_{12|h+1}^{A_1}\D_{12}^{2h}.
\end{equation}
The exponent $2h$ of the scalar derivative is not necessarily integer as it is related to the scaling dimensions of the fields.
 
Having defined the derivatives, it is now possible to write the OPE as
\begin{eqnarray}
\Op{i}{1}\Op{j}{2}&=&(\mathcal{T}_{12}^{\boldsymbol{N}_i}\Gamma)(\mathcal{T}_{21}^{\boldsymbol{N}_j}\Gamma)\cdot\sum_k\sum_{a=1}^{N_{ijk}}\frac{\cOPE{a}{i}{j}{k}\tOPE{a}{i}{j}{k}{1}{2}}{\ee{1}{2}{p_{ijk}}}\cdot\nonumber\\
&&\D_{12}^{(d,h_{ijk}-n_a/2,n_a)}(\mathcal{T}_{12\boldsymbol{N}_k}\Gamma)*\Op{k}{2},\label{eq:OPE}
\end{eqnarray}
where $\cOPE{a}{i}{j}{k}$ are arbitrary coefficients, one for each independent structure, while $\tOPE{a}{i}{j}{k}{1}{2}$ is a tensor that contracts Lorentz indices of different objects in the OPE. The exponents $p_{ijk}$ and $h_{ijk}$ are determined by comparing homogeneity of the two sides of the OPE.

The last ingredient of our framework are the half-projectors $\mathcal{T}_{12}^{\boldsymbol{N}_i}\Gamma$ appearing in Eq.~(\ref{eq:OPE}). By construction, these objects are transverse to match the transversality, Eq.~(\ref{eq:transverse}), of the operators $ \Op{i}{1}$ and $\Op{j}{2}$. $\mathcal{T}_{12}^{\boldsymbol{N}_i}\Gamma$ also matches the Lorentz representation of operator $\Op{i}{1}$. A simple example will illustrate this concept. Suppose we consider an operator with $\boldsymbol{N}=(0,1,0,\ldots)$ that is a two-index antisymmetric tensor. In this case, $\mathcal{T}_{12}^{\boldsymbol{N}}\Gamma\propto\Gamma^{AB}\eta_{1A}(\A_{12})_{BC}$. We termed these objects half-projectors because one gets a projection operator by contracting together two $\Gamma$ matrices $\mathcal{P}_{\boldsymbol{N}}\propto\Gamma^{AB}\Gamma_{AB}$, where the spinor indices of the $\Gamma$ matrices are free. For more complicated representations, $\mathcal{T}_{12}^{\boldsymbol{N}_i}\Gamma$ can be constructed recursively, see Ref.~\cite{longpaper}. It will be important shortly, that as far as coordinate dependence is concerned, the half projectors are simply polynomials in the coordinates $\eta^A$ and also contain dot products $\ee{1}{2}{}$.


\section{Applying the OPE}\label{sec:Exton}

Using the OPE in an $M$-point correlation function reduces it to a function with one fewer point. This can only be of practical use if the derivative operator in Eq.~(\ref{eq:OPE}) can be evaluated on the most general function of coordinates that appear in an $M-1$ function. We only need to be concerned with dot products since any variable with free Lorentz indices can be absorbed into derivatives by the identity in Eq.~(\ref{eq:derivative-coordinates}).  Thus, the most general expression we need is
\begin{equation}\label{eq:I}
I_{ij}^{(d,h,n;\boldsymbol{p})A_1\cdots A_n}=\D_{ij}^{(d,h,n)A_1\cdots A_n}\prod_{a\neq i,j}\frac{1}{\ee{j}{a}{p_a}},
\end{equation}
where the derivative operator $\D_{ij}$ is identical to $\D_{12}$ defined in the previous section, except we replaced $\eta_1$ with $\eta_i$ and $\eta_2$ with $\eta_j$ as we are dealing with multiple coordinates.

The natural variables for conformal blocks are the invariant cross-ratios. For $M>3$ we single out two coordinates, $\eta_k$ and $\eta_l$. Our basis for the cross-ratios is
\begin{eqnarray}
x_a&=&\frac{\ee{i}{j}{}\ee{k}{\ell}{}\ee{i}{a}{}}{\ee{i}{k}{}\ee{i}{\ell}{}\ee{j}{a}{}},\\
z_{ab}&=&\frac{\ee{i}{k}{}\ee{i}{\ell}{}\ee{a}{b}{}}{\ee{k}{\ell}{}\ee{i}{a}{}\ee{i}{b}{}},
\end{eqnarray}
where $a,b\neq i,j$. For convenience, we also define a homogeneous derivative
\begin{equation} 
\bar{\D}_{ij;k\ell|h}^A=\frac{\ee{i}{j}{\frac{1}{2}}\ee{k}{\ell}{\frac{1}{2}}}{\ee{i}{k}{\frac{1}{2}}\ee{i}{\ell}{\frac{1}{2}}}\D_{ij|h}^A.
\end{equation}
In terms of $\bar{\D}$ we define
\begin{equation}\label{eq:Ibar}
\bar{I}_{ij;k\ell}^{(d,h,n;\boldsymbol{p})}=\bar{\D}_{ij;k\ell}^{(d,h,n)}\prod_{a\neq i,j}x_a^{p_a}.
\end{equation}
By definition, $\bar{I}_{ij;k\ell}^{(d,h,n;\boldsymbol{p})}$ is homogeneous in every coordinate and it is proportional to $I_{ij}^{(d,h,n;\boldsymbol{p})}$ in Eq.~(\ref{eq:I}).

Expression for $\bar{I}_{ij;k\ell}^{(d,h,n;\boldsymbol{p})}$ is the central result here. It was obtained mostly by recursion. In the following, we use $(\alpha)_\beta=\Gamma(\alpha+\beta)/\Gamma(\alpha)$ to denote the Pochhammer symbol. We also singled out one of the cross-ratios $x_m$ and traded the remaining $x_a$'s for $y_a=1-x_m/x_a$ when $a\neq i,j,m$.
\begin{widetext}
\begin{eqnarray}
\bar{I}_{ij;k\ell}^{(d,h,n;\boldsymbol{p})}&=&(-2)^h(\bar{p})_h(\bar{p}+1-d/2)_hx_m^{\bar{p}+h}\sum_{\substack{\{q_r\}\geq0\\\bar{q}=n}}S_{(\boldsymbol{q})}x_m^{\bar{q}-q_0-q_i}K_{ij;k\ell;m}^{(d,h;\boldsymbol{p};\boldsymbol{q})}(x_m;\boldsymbol{y};\textbf{z}),\label{eq:master}\\
S_{(\boldsymbol{q})}^{A_1\cdots A_{\bar{q}}}&=&g^{(A_1A_2}\cdots g^{A_{2q_0-1}A_{2q_0}}\bar{\eta}_1^{A_{2q_0+1}}\cdots\bar{\eta}_1^{A_{2q_0+q_1}}\cdots\bar{\eta}_N^{A_{\bar{q}-q_N+1}}\cdots\bar{\eta}_N^{A_{\bar{q}})},\nonumber\\
K_{ij;k\ell;m}^{(d,h;\boldsymbol{p};\boldsymbol{q})}&=&\frac{(-1)^{\bar{q}-q_0-q_i-q_j}(-2)^{\bar{q}-q_0}\bar{q}!}{\prod_{r\geq0}q_r!}\frac{(-h-\bar{q})_{\bar{q}-q_0-q_j}(p_m)_{q_m}(\bar{p}+h)_{\bar{q}-q_0-q_i}}{(\bar{p})_{\bar{q}-2q_0-q_i-q_j}(\bar{p}+1-d/2)_{-q_0-q_i-q_j}}\prod_{a\neq i,j,m}(p_a)_{q_a} K_{ij;k\ell;m}^{(d+2\bar{q}-2q_0,h+q_0+q_j;\boldsymbol{p}+\boldsymbol{q})},\nonumber\\
K_{ij;k\ell;m}^{(d,h;\boldsymbol{p})}&=&\!\!\!\sum_{n_a,n_{am},n_{ab}\geq0}\frac{(-h)_{\bar{n}_m+\bar{\bar{n}}}(p_m)_{\bar{n}_m}(\bar{p}+h)_{\bar{n}-\bar{\bar{n}}}}{(\bar{p})_{\bar{n}+\bar{n}_m}(\bar{p}+1-d/2)_{\bar{n}_m+\bar{\bar{n}}}}\prod_{a\neq i,j,m}\frac{(p_a)_{n_a} y_a^{n_a} x_m^{n_{am}} z_{am}^{n_{am}} }{n_{am}!(n_a-n_{am}-\bar{n}_a)! y_a^{n_{am}}}\prod_{\substack{a,b\neq i,j,m\\b>a}}\frac{1}{n_{ab}!}\left(\frac{x_mz_{ab}}{y_ay_b}\right)^{n_{ab}}.\nonumber
\end{eqnarray}
\end{widetext}
Further details can be found in Ref.~\cite{longpaper}. It is clear that the expression for $\bar{I}_{ij;k\ell}^{(d,h,n;\boldsymbol{p})}$ is fairly complicated. However, it is the most general function needed to construct any conformal block with $M$ points.

For $M=4$, $I_{ij}^{(d,h,n;\boldsymbol{p})}$ in Eq.~(\ref{eq:I}) is directly related to conformal blocks. For $M>4$, $I_{ij}^{(d,h,n;\boldsymbol{p})}$ can be used to build the blocks. Function $K_{ij;k\ell;m}^{(d,h;\boldsymbol{p})}$, defined in the last line of Eq.~(\ref{eq:master}), is precisely the Exton $G$-function when $M=4$~\cite{Exton}. For $M>4$, $K_{ij;k\ell;m}^{(d,h;\boldsymbol{p})}$ generalizes the Exton function to $M$ points.


\section{Discussion}\label{sec:discussion}

There are several advantages of the formalism described in this article for computing conformal blocks. Every Lorentz representation appears on the same footing: there is no significant distinction between bosonic and fermionic operators. All operators carry only spinor indices, with an even number of indices for bosons and an odd number for fermions. The particulars of the representation are encoded in the half-projectors $\mathcal{T}_{12}^{\boldsymbol{N}_i}\Gamma$ introduced in Eq.~(\ref{eq:OPE}). The half-projectors are functions of coordinates and the $\Gamma$ matrices of $SO(d+2)$. The half-projectors are straightforward to write for any fundamental representation. Half-projectors for larger representations can be constructed recursively starting from the smaller ones.

Given the explicit form for $\bar{I}_{ij;k\ell}^{(d,h,n;\boldsymbol{p})}$ in Eq.~(\ref{eq:Ibar}), no evaluations are needed to obtain a conformal block. There is no need to solve differential equations or to compute integrals. The problem has been reduced to putting together Lego bricks. The half-projectors and $\bar{I}_{ij;k\ell}^{(d,h,n;\boldsymbol{p})}$ form a complete Lego set for conformal blocks. We are not implying that obtaining blocks with large Lorentz representations, or many points, is trivial, but that our formalism offers a prescription how to do that and provides all the necessary ingredients.

$\bar{I}_{ij;k\ell}^{(d,h,n;\boldsymbol{p})}$ is very complicated. Part of that complication stems from its generality. It can be used for $M$-point functions, not just four-point functions one might be most interested in. There are only two cross-ratios for four points. Hence, the set of cross-ratios $x_m;\boldsymbol{y};\textbf{z}$, in  Eq.~(\ref{eq:Ibar}), reduces to just two: one $x$ one $y$, and no $z$'s. We computed the most general function because the methods that yielded the answer for four points were sufficient to extend the answer to $M$ points and the corresponding $M(M-3)/2$ cross-ratios.

The OPE plays the primary role in our approach. It extends $(M-1)$-point functions to $M$-point functions. Focusing on four-point functions alone, one could use the OPE only once since three-point functions can be written relatively easily. It is, however, possible to construct every possible correlator using the OPE starting with the two-point function. The two-point function follows the OPE of two operators with the identity operator on the right-hand side. Moving on to three points, one can construct a basis for the three-point functions and relate them to the OPE coefficients. The conformal blocks at four, and more, points are the next steps in employing the OPE. 

The functions $ K_{ij;k\ell;m}^{(d,h;\boldsymbol{p};\boldsymbol{q})}$ and its little sibling $K_{ij;k\ell;m}^{(d,h;\boldsymbol{p})}$ in Eq.~(\ref{eq:Ibar}) have a number of interesting properties~\cite{longpaper}. For example, contiguous relations express $K_{ij;k\ell;m}^{(d,h;\boldsymbol{p})}$ in terms of $K_{ij;k\ell;m}^{(d+2,h;\boldsymbol{p}')}$. Such relations suggest associations between conformal blocks that we are only starting to explore.

We hope that our methods will lead to further advances in conformal bootstrap. The numerical bootstrap can benefit from derivations of previously unknown conformal blocks. It might also be possible to formulate analytic bootstrap completely in the embedding space, which seems more natural for CFTs compared to the position space.


\section*{Acknowledgments}

We thank W.~Goldberger, W.~Ma, and V.~Prilepina for helpful discussions. The work of JFF is supported by NSERC and FRQNT.

\end{document}